\documentclass[preprint,review,12pt]{elsarticle}

\usepackage[T1]{fontenc}
\usepackage{lmodern}
\usepackage{pbox}
\usepackage{array}
\usepackage{makecell}
\usepackage{multirow}
\usepackage{dsfont}
\usepackage{url}
\usepackage{subfig}
\usepackage{amsmath,amsfonts,amssymb, amsthm,bm}
\usepackage{graphicx}
\usepackage{float}
\usepackage{booktabs,tabularx}

\usepackage{stfloats}
\usepackage[toc,page]{appendix}

\usepackage{pbox}
\usepackage{array}
\usepackage{lineno}

\begin{document}

\begin{frontmatter}

\title{Are Bitcoins price predictable? Evidence from machine learning techniques using technical indicators.}

 \author[label1]
 {Samuel Asante Gyamerah}
 \address[label1]{Pan African University, Institute for Basic Sciences, Technology, and Innovation, Kenya}
 \tnotetext[label1]{correspondence:\,\, saasgyam@gmail.com}

\begin{abstract}
The uncertainties in future Bitcoin price make it difficult to accurately predict the price of Bitcoin. Accurately predicting the price for Bitcoin is therefore important for decision-making process of investors and market players in the cryptocurrency market. Using historical data from 01/01/2012 to 16/08/2019, machine learning techniques (Generalized linear model via penalized maximum likelihood, random forest, support vector regression with linear kernel, and stacking ensemble) were used to forecast the price of Bitcoin. The prediction models employed key and high dimensional technical indicators as the predictors. The performance of these techniques were evaluated using mean absolute percentage error (MAPE), root mean square error (RMSE), mean absolute error (MAE), and coefficient of determination (R-squared). The performance metrics revealed that the stacking ensemble model with two base learner (random forest and generalized linear model via penalized maximum likelihood) and support vector regression with linear kernel as meta-learner was the optimal model for forecasting Bitcoin price. The MAPE, RMSE, MAE, and R-squared values for the stacking ensemble model were $0.0191\%$, $15.5331$ USD, $124.5508$ USD, and $0.9967$ respectively. These values show a high degree of reliability in predicting the price of Bitcoin using the stacking ensemble model. Accurately predicting the future price of Bitcoin will yield significant returns for investors and market players in the cryptocurrency market.
\end{abstract}

\begin{keyword}
 Bitcoin volatility \sep Machine learning \sep stacking ensemble \sep Bitcoin price forecasting \sep technical indicators
\end{keyword}

\end{frontmatter}
%\linenumbers

%\linenumbers

\section{Introduction}
\noindent 
Bitcoin is considered as the world's largest digital currency by market  capitalisation{\footnote{estimated as \$182,675,714,614}} \cite{bitcoin}. Bitcoin has generated a lot of returns for market players and investors alike{\footnote{\url{https://www.investopedia.com/articles/investing/123015/if-you-had-purchased-100-bitcoins-2011.asp}}}. Nevertheless, there is a strong fluctuations in the price of Bitcoin \cite{wolla2018bitcoin} leading to price uncertainties; a situation that threatens its potential to function as a currency. Bitcoin is therefore seen as a highly volatile currency. Market players and analysts have associated different factors to the high price volatility of Bitcoin. Among these factors are: a relatively small market as compared to   traditional assets such as fiat currencies, bonds, and stocks, low liquidity which increases price fluctuations, regulation problems and failure, news events, shifting sentiments, and high speculations. The volatile nature of Bitcoin makes price prediction very difficult for most investors and market players. Hence, we develop machine learning predicting models that can accurately forecast the price of Bitcoin to help investors and players in the Bitcoin market. In this study, Bitcoin is used for the prediction problem because of the magnitude of its market capitalization. 
\\
[2mm]
Analogous to stock price and foreign exchange prediction using machine learning algorithms, the price of Bitcoin can also be fpredicted using different machine learning techniques. However, literatures on Bitcoin price prediction using machine learning techniques are not exhaustive. \cite{greaves2015using} analyzed the prediction strength of blockchain
network-based features on Bitcoin's future price. The classification accuracy for their prediction was about 55\%. As indicated by \cite{gyamerah2019stock}, an accuracy value closer or less than 50\% for a binary classification problem is as good as randomly selecting the labels. Hence, the blockchain network-based algorithm they employed was not effective in predicting the movement in the price of Bitcoin. In their study, \cite{jang2017empirical} used Bayesian neural networks (BNNs), linear and support vector regression models to predict the price of Bitcoin. BNN performed better in predicting the price of Bitcoin as compared to linear and support vector regression models. Using a genetic algorithm based selective neural network, \cite{sin2017bitcoin} studied the relationship between the predictors of Bitcoin and the day-ahead change in Bitcoin price. The model was later used to predict the day-ahead movement of Bitcoin price. By implementing a Bayesian optimised recurrent neural network 
and a Long Short Term Memory network on Bitcoin price time series data obtained from Bitcoin Price Index, \cite{mcnally2018predicting} identified the percentage accuracy for which the price of Bitcoin in United States Dollars (USD) can be predicted. They compared the deep learning models to an autoregressive integrated moving average (ARIMA) model and concluded that deep learning models are better in classification prediction than the ARIMA model. 
\\
[2mm]
From the Bitcoin prediction literatures, there have not yet been empirical studies using key and high dimensional technical indicators as features for Bitcoin price prediction. Also the selected individual and stacking algorithms have not been explored in literature. It is therefore worthwhile to apply these algorithms using 34 key technical indicators for Bitcoin price predictions. The general objective of this paper is to determine the accuracy of predicting the price of Bitcoin in the midst of price uncertainties. The specific contributions are: 1) to build an accurate prediction model that incorporates key and high dimensional technical indicators on the cryptocurrency market 2) to predict the price of Bitcoin using Generalized linear model via penalized maximum likelihood, random forest, support vector regression with linear kernel, 3) to compare these individual machine learning models to Stacking ensemble model, 4) to add to the scarce empirical evidence in predicting the price of Bitcoin using machine learning techniques reported in literature.  
\\
[2mm]
The rest of the paper is organized as follows: section \ref{ml_techniques} provides explanation for the machine learning techniques used in the study; the data, technical indicators, data pre-processing, and evaluation metrics used for the study are presented in section \ref{methodology}; section \ref{results_discussions} describes the empirical results and analysis of the prediction models; and the conclusions are outlined in section \ref{conclusion}.

\section{Machine Learning Forecasting Techniques}
\label{ml_techniques}
\noindent
In this study, we use machine learning as a tool for forecasting the price of Bitcoin. The choice of an optimal machine learning algorithm for forecasting is a major factor to consider in any forecasting problem. For this reason, the chosen machine learning technique should be able to forecast the price of Bitcoin with a small margin of error. 

\subsection{Support Vector Regression (SVR)}
A generalized version of support vector machine (SVM) called the support vector regression (SVR) was proposed by \cite{drucker1997support} in 1996. The output model of SVR relies solely on a subsample of training data. The cost function for constructing the SVR model does not take into consideration any training data that is near to the model prediction. SVR also uses kernels and has demonstrated to be a functional and versatile tool in most real-valued function computation. The following steps can be used to implement SVR.
\\[1mm]
$\mathit{Step 1}$.  Given a training dataset $\{(x_1,y_1), (x_2,y_2), \cdots, (x_i,y_i)\} \subset K \times \mathbb{R}$, where $K$ is a high dimensional space of the input pattern $(K = \mathbb{R}^d)$.
\\[1mm]
$\mathit{Step 2}$. A nonlinear (NL) regression problem can be changed into a functional linear regression problem in $K$ by making use of a linear function called the SVR function, 
\begin{equation}
h(\mathbf{x}) = \mathbf{v}^{T}\cdot\tau(\mathbf{x}) + b,  \qquad\qquad v \in K,\quad b \in \mathbb{R}
\end{equation} 
$h(x)$ is the forecasted Bitcoin price values, the coefficients $\mathbf{v}$ and $b$ can be tuned.
\\[1mm]
$\mathit{Step 3}$. The observed risk, $R(h)$ can be determine as, 
\begin{equation}
R(h) = \frac{1}{N} \sum_{i=1}^{N} \psi_{\epsilon} (y_i, h(x)), 
\label{observed_risk}
\end{equation}
$\psi_{\epsilon} (y_i, h(x))$ represents a $\epsilon$-intensive loss function defined as,
\begin{equation}
\psi_{\epsilon} (y_i, h(x)) = \begin{cases}
|h(\mathbf{x}) - \mathbf{y}| - \epsilon, \qquad \mbox{if} \quad |h(\mathbf{x}) - \mathbf{y}| \ge \epsilon,
\\
0, \qquad\qquad\qquad\qquad \mbox{otherwise} .
\end{cases}
\end{equation}
The purpose of the $\epsilon$-intensive loss function is to restrict the way the model are generalized. 
\\[1mm]
$\mathit{Step 3}$. Using a quadratic optimization problem with inequality constraints, the errors between the training data and the the $\epsilon$-intensive loss function can be estimated, 
\begin{equation}
\begin{aligned}
& {\text{minimize}}
& &\dfrac{1}{2} \| v \|^2 + \lambda \sum_{i=1}^{N} (\vartheta_i + \vartheta_i^*)\\
& \text{subject to}
& & \begin{cases}
y_i - \langle v,x_i \rangle - b \le \epsilon + \vartheta_i, \quad\quad i=1,2,\cdots, N,
\\
\langle v,x_i \rangle + b - y_i \le \epsilon + \vartheta_i^* \quad\quad i=1,2,\cdots, N,
\\
\vartheta_i, \vartheta_i^* \qquad\qquad\qquad\qquad\qquad i=1,2,\cdots, N.
\end{cases}
\end{aligned}
\label{SVR.2}
\end{equation}
$\lambda > 0$ is a constant and it controls the trade-off between the allowable magnitude of the deviation of $\epsilon$ and the flatness of $h$. While the first part of the objective function penalizes large weights, regularize the size of the weight, and preserve the flatness in the regression function, the second part penalizes the training errors associated with $h(\mathbf{x})$ and $\mathbf{y}$.  However, some errors can be allowed by introducing slack variables $\vartheta_i, \vartheta_i^*$ to deal with the infeasible constraints.
\\[1mm]
$\mathit{Step 4}$. By solving equation \ref{SVR.2}, $\mathbf{v}$ can be estimated as, 
\begin{equation}
\mathbf{v} = \sum_{i=1}^{N} (\alpha_i^* - \alpha_i) \tau(\mathbf{x}_i), 
\end{equation}
$\alpha_i^*$, $\alpha_i$ are the Lagrangian multipliers. 
\\[1mm]
$\mathit{Step 3}$ The SVR function is set up as, 
\begin{equation}
\begin{aligned}
h(\mathbf{x}) &= \sum_{i=1}^{N} (\alpha_i^* - \alpha_i) K(\mathbf{x}_i, \mathbf{x}_j) + b,\\
K(\mathbf{x}_i, \mathbf{x}_j) &= e^{-\kappa \| \mathbf{x}_i - \mathbf{x}_j \|^2}, \qquad \kappa > 0,
\end{aligned}
\end{equation}
$K(\cdot)$ is a Kernel function. 
\\
[2mm]
Generally, the performance of SVR depends on the settings of the global parameters: Cost ($C$) controls the trade-off in the model complexity and extent to which the variance greater than $\epsilon$ can allowed, $\epsilon$ controls the width of the insensitive areas, and the Kernel function (K). Selecting an optimal value for these parameters is complicated since SVR depends on all the three parameters.

\subsection{Random Forest (RF)}
Random forest is an ensemble approach based on the idea that ensemble of weak learners (decision trees) when combined would result in a strong learner \cite{breiman2017classification,breiman2001random}. Using Breiman's bagger, each of the variables is considered in every split. Due to the principle of Strong Law of Large Numbers, over-fitting is not a problem in random forest. For this reason, RF always converges. The strength of each single-tree classifier and a measure of their dependencies contributes to the accuracy of random forest. For implementation of the random forest algorithm, the interested reader should see \cite{breiman2001random}.
\\
[2mm]
For optimal performance of random forest model, the number of trees (ntree) and the number of variables sampled as candidates for each split (mtry) must be carefully selected. For regression problems, mtry = $\frac{n}{3}$ (where $n$=number of features used for the prediction). The fraction of the training data that is randomly selected to suggest the next tree in the expansion is called the subsampling fraction or the bag.fration. The default value of bag.fraction is 0.5. However, this value can be increase if the training sample is small.

\subsection{Generalized linear model via penalized maximum likelihood (GLMNET)}
\noindent
Generalized linear model via penalized maximum likelihood is a highly robust method for fitting the entire lasso or elastic-net regularization path for linear regression \cite{friedman2016lasso}. 
GLMNET can take advantage of the sparsity in the features. It can fit linear, multi-response linear, multinomial, logistic, and poisson regression models. Different predictions can be obtained from the fitted regression models.
GLMNET solves the following problem
 \begin{equation}
 \min\limits_{\alpha_0,\alpha} \dfrac{1}{N} \sum_{i=1}^{N}w_i L(y_i, \alpha_0 + \alpha^Tx_i) + \lambda[(1-\gamma)||\alpha||_2^{2}/2 + \gamma||\alpha||_1], 
 \label{glmnet}
 \end{equation}
for a grid of values of $\lambda$ for the full bounds. $L(y, \vartheta)$ is defined as the negative log-likelihood contribution for data point $i$. The elastic-net penalty is controlled by $\gamma$, and connects the gap between lasso ($\gamma=1$) and ridge ($\gamma=0$) penalty. The tuning parameter $\lambda$ regulates the general strength of the penalty. The ridge penalty reduces the coefficients of correlated features towards each other. The lasso penalty hand pick one of the features and drop the other remaining features. The elastic-net penalty combines the ridge and lasso penalty; if features are correlated in groups, a $\gamma=0.5$ is likely to select the groups in or out simultaneously.

\subsection{Stacking Ensemble Leaner}
\begin{figure}[H]
	\centering
	\includegraphics[height=12cm,width=14cm]{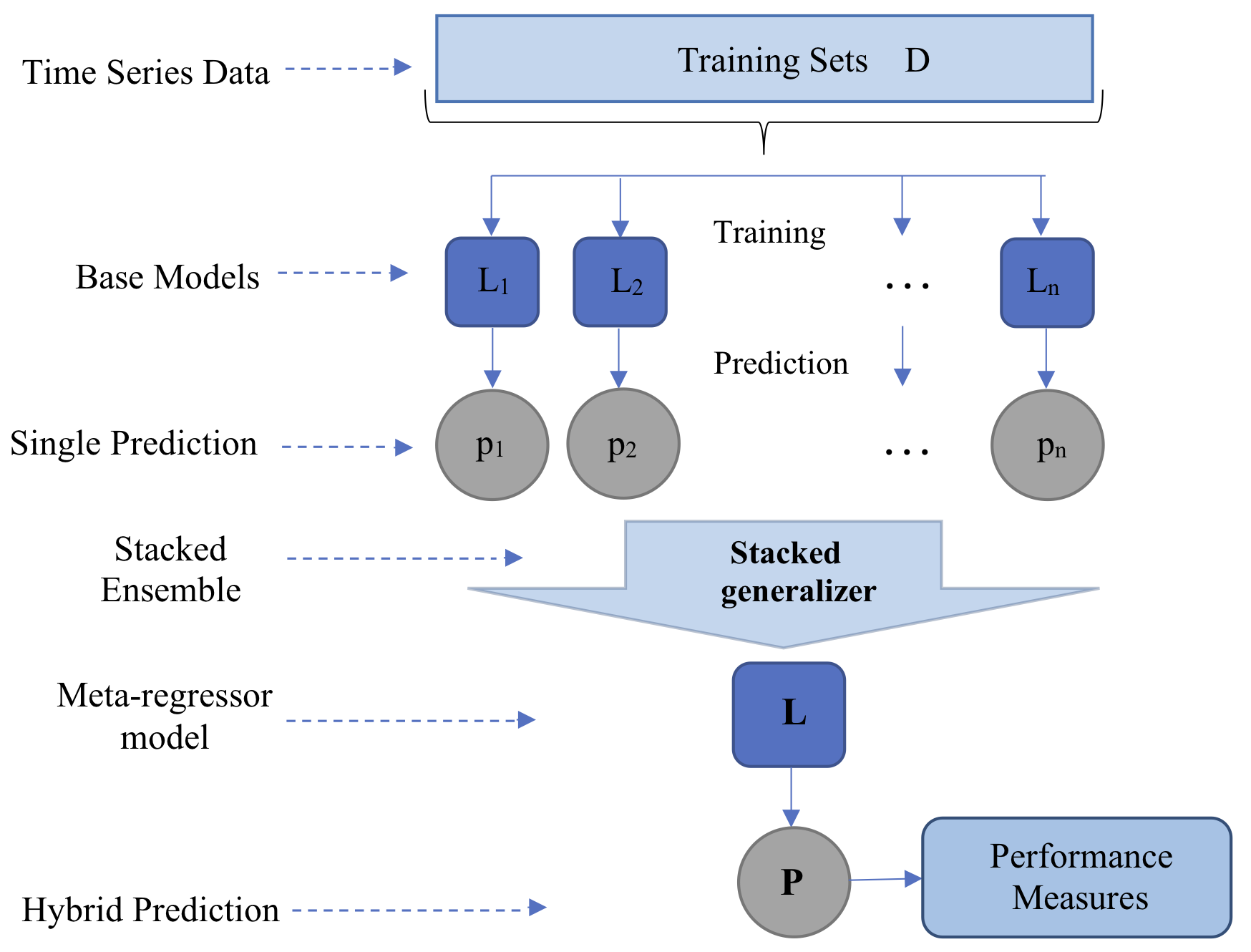}
	\caption{A graphical representation of stacking ensemble scheme.}	
	\label{Stack_algo}
\end{figure}

\noindent
Ensemble learning is a machine learning meta-algorithms where ``weak learners'' are trained and combined into one predictive model to reduce bias (boosting), variance (bagging), or increase the accuracy of predictions (stacking). The concept of ensemble methods is that when weak learners are rightly combined, the resulting model is robust as compared to the individual weak learners. Stacking ensemble is less widely used than boosting and bagging \cite{sewell2008ensemble}. In contrast to boosting and bagging, stacking may be used to combine models of different types. In stacking ensemble, a new model from a meta-regressor learns how to optimally combine the predictions of other existing models from weak learners. That is, the base level weak models (made up of different learning algorithms) are trained on the training dataset and a meta-model is trained using the outputs of the base level model as features. Hence, stacking ensemble learning method can be considered as a ``heterogeneous ensemble model''. From literatures \cite{gyamerah2019stock,tsai2009stock}, predictive models based on stacking ensemble models are usually better than individual model. Figure \ref{Stack_algo} is the visual diagram of stacking ensemble scheme.

\section{Methodology}
\label{methodology}

\subsection{Data}
\label{data}
\begin{figure}[H]
	\centering
	\includegraphics[height=06cm,width=14cm]{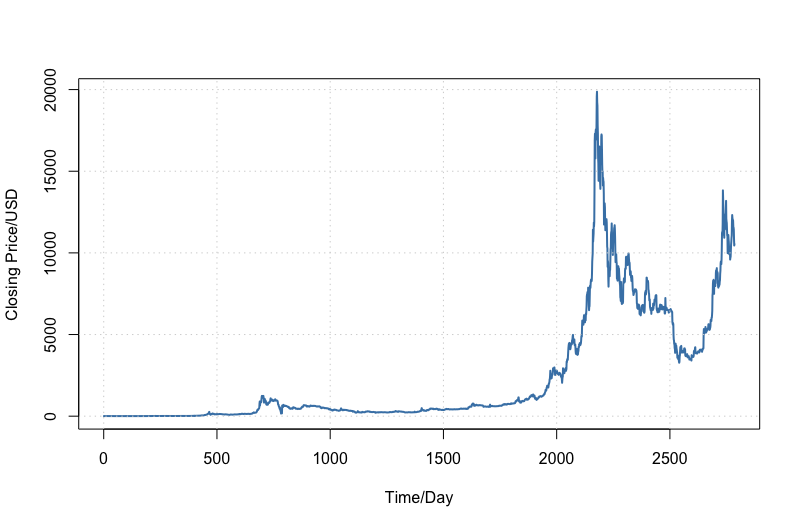}
	\caption{Bitcoin closing price}	
	\label{bitcoin}
\end{figure}

\begin{figure}[H]
	\centering
	\includegraphics[height=06cm,width=14cm]{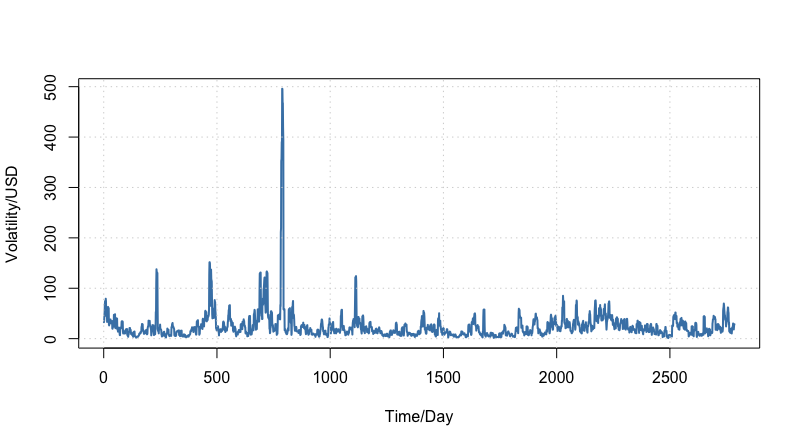}
	\caption{Volatility of Bitcoin}	
	\label{volatility}
\end{figure}

\noindent
Daily dataset of Bitcoin prices and other indicators (High, Low, Open, Volume, SMA5, SMA13, SMA20	SMA30, SMA50, EMA5, EMA12, EMA26, EMA50, MACDLine, MACDSignalLine, MACDHistogram, SMABollBands5, BBands5Up, BBands5Down, SMABollBands13, BBands13Up, BBands13Down, SMABollBands20, BBands20Up, BBands20Down, Volatility) were taken from the CryptoCompare website {\footnote{\url{https://www.cryptocompare.com/}}}. The daily dataset expanded from $01/01/2012$ to $16/08/2019$ making a total of $2785$ trading days. The dataset was divided into training ($01/01/2012$ to $05/02/2018$) and testing ($06/02/2018$ to $16/08/2019$) data. The total sample size of the training and testing data were $2228$ and $557$ respectively. Daily dataset was taken because most investors make decisions to buy or sell a share of a Bitcoin based on the daily closing price of the market. The training data was used in training the machine learning models and the testing data was used in evaluating the performance of the models.
\\
The closing price was used as the general measure of Bitcoin price for the sample period under study. Figure \ref{bitcoin} and \ref{volatility} presents the Bitcoin price dynamics  and the volatility in Bitcoin price over the selected days under study. Clearly, the price of Bitcoin is highly volatile as stated in the introduction section.

\begin{table}[H]
\caption{Descriptive statistics of the Bitcoin dataset from 01/01/2012 to 16/08/2019}
	\centering
%	\resizebox{\columnwidth}{!}{%
		\begin{tabular}{ccccccc}
			\hline
	& Min  & Max & Mean & Std. Dev.  \\
			\hline
Close & 4.22 & 19345.49  & 2259.77 & 3422.389   \\
High  & 4.40 &  19870.62 & 2330.46 &  3547.90 \\
Low  & 3.88 & 18750.91 & 2173.56 & 3264.26 \\
Open  & 4.22 &  19346.60 & 2256.02 & 3419.13   \\
			\hline
		\end{tabular}
%	}
	\label{descriptive_statistics}
\end{table}

\subsubsection{Technical indicators}
\label{indicators}
\noindent
Technical analysis of a cryptocurrency is founded on the assumption that all the important information about a specific cryptocurrency is incorporated in its price and/or other market data like the Volume traded. That is, the dynamics of the historical price and other market data control the decision of market players and investors in the cryptocurrency market. Technical indicators are important tools that can be used to transform price patterns into actionable trading plans. They can therefore be used as features to predict future prices. By applying simple but relevant rules to historical price data, different technical indicators can be generated. The objective of a technical indicator in a cryptocurrency market is to analyze trends in the price of a cryptocurrency in order to forecast the future price of the cryptocurrency. Below are the technical indicators that were calculated from the extracted Bitcoin time series data. These indicators are transformed to features for the forecasting models.
\noindent
\\
[1mm]
Simple Moving Average (SMA): A type of moving average that computes the arithmetic average price over a specific period. 
\begin{equation}
SMA_w = {\sum_{i=0}^{w-1} C_{d-i}}{w},
\label{sma}
\end{equation}
where $C_d$ is the closing price for day $d$ and $w$ is a window size. Simple moving average of order 5 (SMA5), order 13 (SMA13), order 20 (SMA20), order 30 (SMA30) and order 50 (SMA50) were extracted from the cryptocompare website. 
\\
[1mm]
Exponential Moving Average (EMA): A moving average where the weights of historical prices decreases exponentially. It calculates an exponentially-weighted mean, giving more weight to current observations. EMA of order 5, 12, 26, and 50 denoted as EMA5, EMA12, EMA26, and EMA50 were extracted from cryptocompare website. 
\begin{equation}
EMA_t = \dfrac{\sum_{i=0}^{w-1}C_{d-i}}{w}
\label{ema}
\end{equation}
\\
[1mm]
Weighted Moving Average (WMA): WMA is similar to an EMA, but with linear weighting if the length of weights is equal to $w$.
\\
[1mm]
Average True Range (ATR): It measures the volatility of a High-Low-Close series. 
\begin{equation}
ATR_w = EMA_w \Big(max (H_d - L_d, abs(H_d - C_{d-1}),abs(L_d - C_{d-1}))\Big),
\label{atr}
\end{equation}
where $H_d, L_d,$ and $C_d$ are the price high, price low, and closing price at day $d$ respectively. 
\\
[1mm]
Chaikin Accumulation/Distribution line (AD): It measures the money flowing into or out of a Bitcoin market. 
\begin{equation}
AD = AD_{d-1} + \dfrac{(C_d - L_d)-(H_d - C_d)}{H_d - L_d}V_d,
\label{ad}
\end{equation}
where $V_d$ is the volume traded at day $d$. 
\\
[1mm]
Commodity Channel Index (CCI): It identifies cyclical turns in Bitcoin price. CCI can be used to evalaute whether a bitcoin is overbought or oversold.   
\begin{equation}
CCI_w = \dfrac{\varSigma_d - SMA_w(\varSigma_d)}{ 0.015 \sum_{i=1}^{w}|\varSigma_{d-i+1} - SMA_w(\varSigma_d)| / w }, 
\label{cci}
\end{equation}
$\varSigma_d = H_d + L_d + C_d$
\\
[1mm]
Rate of change (ROC): It calculates the rate of change relative to the Bitcoin closing prices over a period of time. 
\begin{equation}
ROC_w = \dfrac{C_d - C_{d-w}}{C_{d-w}}
\label{roc}
\end{equation}
\\
[1mm]
Momentum (MOM): It measures the change in price relative to the actual price levels.
\begin{equation}
MOM_w = C_d - C_{w-1}
\label{mom}
\end{equation}
\\
[1mm]
Moving Average Convergence Divergence (MACD): It is the most popular and widely used technical indicator. It uses the moving averages to determine the momentum of a cryptocurrency. The three components of MACD: MACD signal, signal line, and histogram were calculated. MACD line is calculated as the difference between 12 period EMA and the 26 period EMA. The MACD signal line is a 9 period EMA of the MACD line and the MACD histogram is the difference between the MACD line and the MACD signal line. MACDLine, MACDSignalLine, and MACDHistogram were obtained from the website of cryptocompare. 
\begin{equation}
\begin{aligned}
MACD line &=\mbox{12d EMA} - \mbox{26d EMA}
\\
MACD signal line &=\mbox{9d EMA of MACD line}
\\
MACD histogram &=\mbox{MACD line} - \mbox{signal line}
\end{aligned}
\label{mean}
\end{equation}
\\
[1mm]
Bollinger Band (BBands/BollBands): It is a method used to compare a cryptocurrency volatility and price levels over a period of time. The upper (Up) and lower (Down) BBands were also calculated. The upper and lower BBands are calculated as the standard deviations above and below the moving average.
\\
[1mm]
Stochastic Oscillator (stochOSC): A momentum indicator that relates the location of each day's closing price relative to the high/low range over the past $n$ periods. 
\begin{equation}
stochOSC = \dfrac{C_d - LL_w}{HH_w - LL_w}
\label{stoc}
\end{equation}
where $LL_w$ and $HH_w$ are respectively the mean lowest low and highest high prices for previous $d$ days.

\subsection{Data pre-processing}
\label{data_preprocessing}
To make the data more relevant for the machine learning forecasting models, the time series data was pre-processed.  

\subsubsection{Data Transformation}
\label{data_transformation} 
\noindent
The Bitcoin time series data (Close, High, Low, and Volume) was transformed into a set of ten (10) additional technical indicators which differs from the technical indicators extracted from the CryptoCompare website. These technical indicators are widely used in financial market literatures and help in price forecasting.

\subsubsection{Data Normalization}
\label{data_normalization}
\noindent
The Bitcoin time series data are converted to the same scale without changing the differences in the range of the price values. The minimum-maximum formula (see equation  \ref{normalization})  was used to normalized the dataset into the range $[0,1]$. Using the anti-normalization equation (see equation \ref{denormalization}), the normalized data points can be changed to the magnitude of the actual data points.
\begin{equation}
x_{normalize} = \dfrac{x - minimum(x)}{maximum(x) - minimum(x)}
\label{normalization}
\end{equation}
\begin{equation}
x = x_{normalize}\big(maximum(x) - minimum(x)\big) + minimum(x)
\label{denormalization}
\end{equation}
where  $maximum(x)$, $minimum(x)$, $x_{normalize}$ are the maximum, minimum value of the inputs and the normalized input value respectively.
\\
$R$ statistical software was used in implementing the data normalization. 

\subsubsection{Feature selection}
\label{feature_selection}
Feature selection is an important step in the Bitcoin forecasting problem. Boruta algorithm was used to select the most important features for the forecasting models. Boruta is a feature ranking and selection machine learning algorithm that uses a wrapper approach buitt on RF algorithm. It iteratively eliminates the features that are less important than random probes. The Boruta package in R \cite{kursa2010feature} was used to select the most important features.

\subsection{Evaluation Metrics}
\label{evaluation_metrics}
\noindent
Equation \ref{rmse}--\ref{r_squared} are the metrics used in evaluating the performance of the forecasting models.
\\
Root mean squared error (RMSE),
\begin{equation}
RMSE = \sqrt{\dfrac{\sum_{i=1}^{N}(A_i - F_i)^2}{N}}
\label{rmse}
\end{equation}
Mean absolute error (MAE),
\begin{equation}
MAE = \dfrac{1}{N}\sum_{i=1}^{N} |A_i - F_i | ,
\label{mae}
\end{equation}
Mean absolute percentage error (MAPE),
\begin{equation}
MAPE = \dfrac{1}{N} \sum_{i=1}^{N} \Big( \dfrac{|A_i - F_i|}{|A_i|}
\Big) \times 100\%,
\label{mape}
\end{equation}
Coefficiet of determination/R-squared ($R^2$)
\begin{equation}
R^2 = 1 - \dfrac{\sum_{i=1}^{n}(A_i - F_i)^2}{\sum_{i=1}^{n}(A_i - \bar{A}_i)^2},
\label{r_squared}
\end{equation}  
\\
where $A_i, \bar{A}_i, F_i$ are the actual, mean, and the forecasted Bitcoin prices. In comparing the techniques, the model that gives a lower RMSE, MAE, and MAPE is considered as the best model with respect to these metrics. A model with a larger R-Squared value is considered to be the best model when using R-Squared as the performance metric. The RMSE, MAE measure ranges from $0$ to $\infty$. MAPE measure (equation \ref{mape}) ranges from $0$ to $100\%$. R-Squared measures the degree of relationship between the forecasted and the real price data and it ranges from $0$ to $1$. In all the machine learning techniques, the testing data was used to evaluate and validate the performance of the model.

\section{Experimental results and discussion}
\label{results_discussions}
\subsection{Feature selection}
\label{results_feature}
\begin{figure}[H]
	\centering
	\includegraphics[height=8cm,width=15cm]{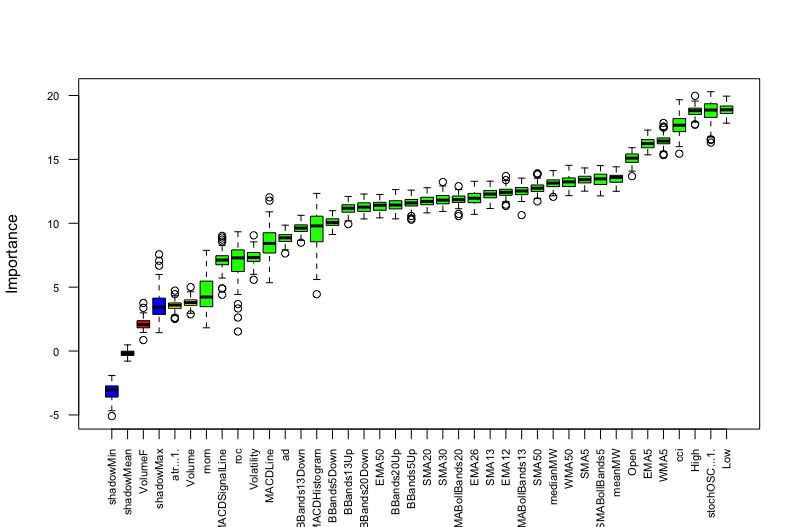}
	\caption{Boruta result plot for technical indicators.}	
	\label{boruta}
\end{figure}

\noindent
Boruta performed 99 iterations in 43.2388 minutes and 34 attributes were confirmed important. One output (volume from (volumeF)) was considered unimportant and two outputs (average true range (atr) and volume to (volume)) were considered to be tentative. Figure \ref{boruta} displays the Boruta result plot for the technical indicators. The plot shows the importance of each of the technical indicators. The columns in green are the `confirmed' technical indicators and the column in red is not. There are two tentative attributes shown in yellow columns. The blue bars (shadowMin, shadowMax) are not technical indicators but are used by Boruta algorithm to determine if an indicator is important or not important. Table \ref{boruta_meanImp} presents the mean importance of the technical indicators from the Boruta algorithm. 

\begin{table}[H]
	\caption{Selected features using Boruta algorithm}
	\centering
	\resizebox{\columnwidth}{!}{%
		\begin{tabular}{cc|cc|cc}
			\hline
			Feature & meanImp  & Feature & MeanImp & Feature & meanImp  \\
			\hline
			Low & 8.6621 & EMA12  & 5.5459 & BBands5Down & 4.5395 \\
			stochOSC & 8.5909 & SMABollBands13  & 5.5384 & MACDline & 4.4343  \\
			High & 8.5353 & SMA13  & 5.5015 & BBands13Down & 4.2226  \\
			cci & 8.0096 & SMA30 & 5.3326 & ad & 3.9548 \\
			WMA5 & 7.3011 & EMA26  & 5.3188 & Volatility & 35332 \\
			EMA5 & 7.3638 & SMABollBands20 & 5.2633 & roc & 3.4091 \\
			Open & 6.8225 & SMA20  & 5.2340 & MACDSignalLine & 3.3965\\
			SMABollBands5 & 6.0390 & BBands5Up  & 5.1564 & mom & 2.7701  \\
			meanMW & 6.0287 & BBands20up  & 5.0809 &  &  \\
			SMA5 & 5.9783 &  BBands13Up & 5.0423 &  &  \\
			WMA50 & 5.9061 & BBands20Down & 5.0072 &  &  \\
			medianMW & 5.8782 & EMA50  & 4.9579 &  &  \\
			SMA50 & 5.7759 & MASCDHistogram & 4.5757 &  &  \\
			\hline
		\end{tabular}
	}
	\label{boruta_meanImp}
\end{table}

\subsection{Forecasting with ML techniques}
\label{results_forecasting}
\begin{figure}[H]
	\centering
	\includegraphics[height=07cm,width=15cm]{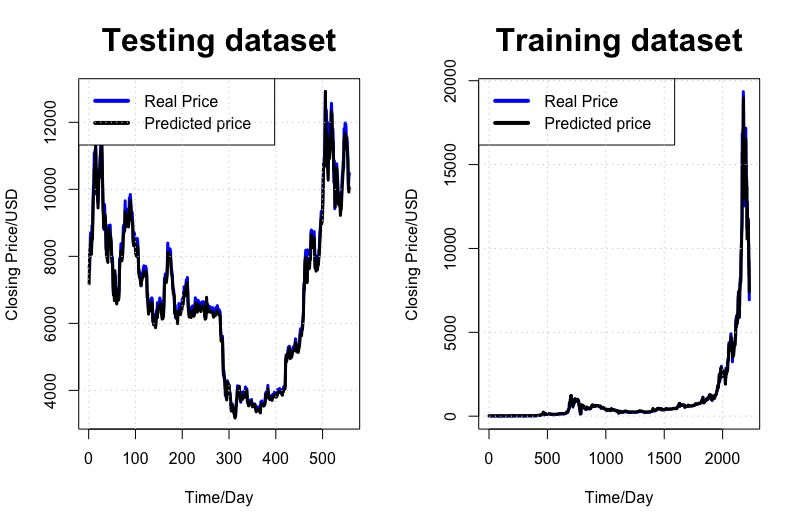}
	\caption{Real and Predicted Bitcoin price of testing and training dataset using generalized linear model via penalized maximum likelihood}	
	\label{glmnet_predictions}
\end{figure}

\begin{figure}[H]
	\centering
	\includegraphics[height=07cm,width=15cm]{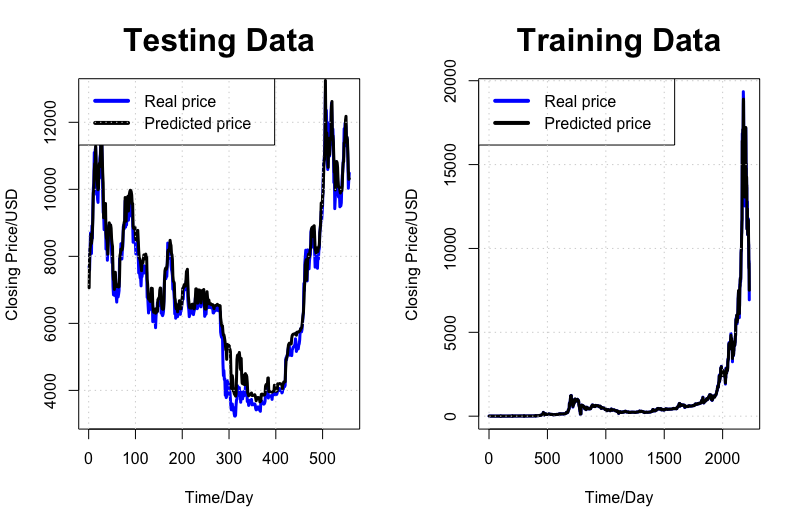}
	\caption{Real and Predicted Bitcoin price of testing and training dataset using random forest}	
	\label{rf_predictions}
\end{figure}

\begin{figure}[H]
	\centering
	\includegraphics[height=07cm,width=15cm]{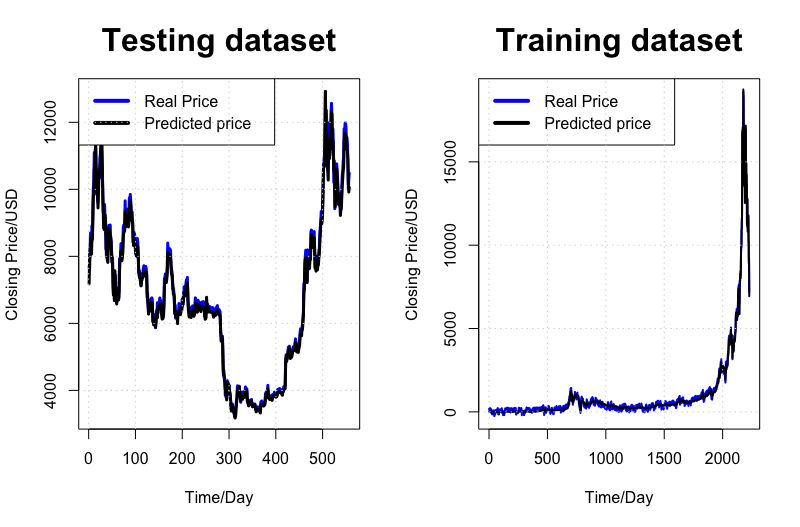}
	\caption{Real and Predicted Bitcoin price of testing and training dataset using support vector regression with linear kernel}	
	\label{esvr_predictions}
\end{figure}

\begin{figure}[H]
	\centering
	\includegraphics[height=07cm,width=15cm]{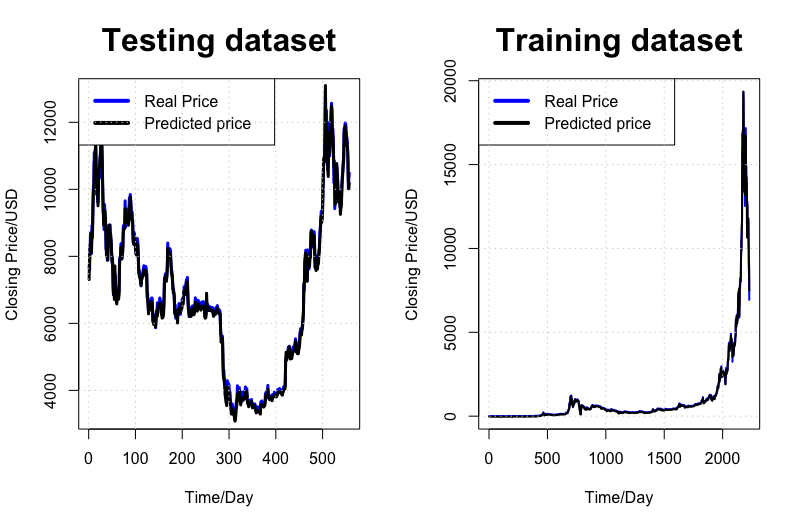}
	\caption{Real and Predicted Bitcoin price of testing and training dataset using stacking ensemble}	
	\label{stack_predictions}
\end{figure}

\begin{figure}[H]
	\centering
\includegraphics[height=06.1cm,width=16cm]{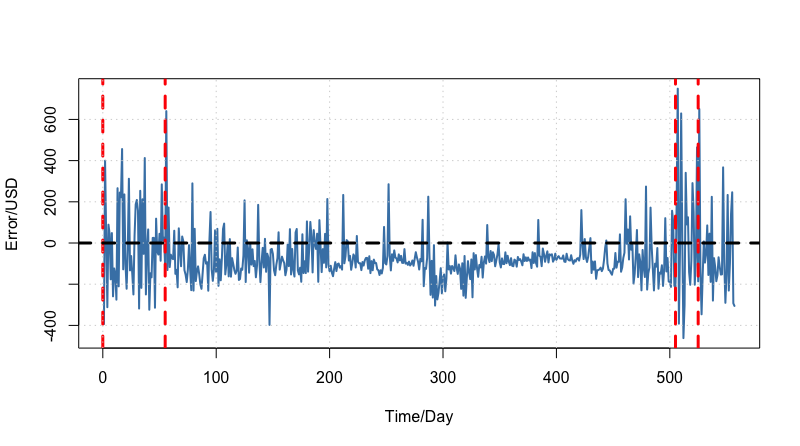}
\caption{Forecasting error of Stacked ensemble model}	
	\label{error_stacking}
\end{figure}

\noindent
Using the training data, the four ML techniques were fine-tuned to select the optimal parameter value for the forecasting model. 
\\
For the generalized linear model via penalized maximum likelihood, resampling was done on a 10 fold cross validation and repeated for 6 times. The smallest root mean square error value was used to select the best model. The final parameters value used to construct the model were $alpha=1$ (pure lasso regression) and $lambda=1 \times 10^{-4}$. Resampling was done on a 12 fold cross validation and repeated for 8 times for random forest algorithm. Using the smallest root mean square error value, the best random forest model was selected for the training model. The final parameter value were ntree=$2500$, mtry=$13$, bag.fraction=$0.75$. Using a 10 fold cross validation, support vector regression was sampled. The final parameter and parameter value used after fine-tuning the model were: svm type=$epsilon$-$regression$, svm-kernel=$linear$ $kernel$, cost=$0.07$, epsilon=$0.1$, number of support vectors=$18$, tolerance=$0.001$. Random forest and generalized linear model via penalized maximum likelihood were used as the base classifier for the stacking ensemble. Resampling for the meta-learner (support vector machine with linear kernel) was repeated 5 times on a 10 fold cross-validation. The tuning parameter of the random forest model `C' was held constant at a value of $1$. 
\\
In all the above ML forecasting models, the CARET package in R \cite{kuhn2017misc} was used for the implementation. 
\\
[2mm]
Evaluation metrics (MAPE, RMSE, MAE, and R-squared) defined in section \ref{methodology} were used to evaluate the performance of the ML algorithms. Table \ref{evaluation_metrics_point} presents the evaluation metrics values for generalized linear model via penalized maximum likelihood, random forest, support vector regression with linear kernel, and stacking ensemble machine learning models using the above parameter values. Evaluation metrics were computed for the forecasted results of training and testing data. 
\\
From the table, stacking ensemble recorded the lowest ($15.5331$ USD) and highest ($0.9967$) mean absolute error and R-squared value respectively for the testing data. Using mean absolute error and R-squared as the performance metrics, the random forest model was the best model for predicting the training data as it recorded values of $10.2114$ USD and $0.9997$ respectively. The mean absolute percentage error ($0.0191\%$) and root mean square error ($15.5331$ USD) values of the stacking ensemble model indicate an optimal performance of the stacking ensemble model in predicting the Bitcoin testing dataset. Random forest recorded the lowest performance metrics in mean absolute percentage error ($0.0063 \%$) and root mean square error ($44.0983$ USD) values for the training dataset. 
\\
Random forest and support vector regression with linear kernel recorded the minimum R-squared values for the testing and training dataset respectively. This indicates that the relationship between the predicted testing values from the random forest model and actual testing data values is not as strong as the other models. The same can be said of the support vector regression with linear kernel for the predicted and actual training data values. Random forest recorded the highest mean absolute percentage error ($0.0548\%$), root mean square error ($398.6882$ USD), and mean absolute error ($305.1938$ USD) for the testing data. Support vector regression with linear kernel recorded the highest absolute percentage error ($3.2798\%$), root mean square error ($139.4931$ USD), and mean absolute error ($121.4614$ USD) for the training data. However, it performed better in predicting the testing data than random forest (see figure \ref{rf_predictions}) and and generalized linear model via penalized maximum likelihood models (see figure \ref{glmnet_predictions}). 
\\
MAPE, RMSE, MAE and R-squared metrics are important measures for evaluating the performance of the models in predicting the price of Bitcoin. It will therefore be bias to use only one of the metrics to select a model for Bitcoin price prediction. From the four performance metrics results in table \ref{evaluation_metrics_point}, the stacking ensemble model was the optimal model. It predicted the testing data with the highest precision and small error values. This is consistent with the visual presentations in Figure \ref{stack_predictions}. Support vector regression with linear kernel model followed closely with higher precision metric values for the testing data. This is evident from the visual plot in figure \ref{esvr_predictions}. As shown in the figure \ref{rf_predictions}, the predictions of random forest on the testing data was not accurate as compared to the other models.
\\
Figure \ref{error_stacking} is the plot of stacking ensemble model point forecasting error for the testing data. From the figure, it is clear that the stacking ensemble model under-forecasted majority of the data points (see the plot below the black dashed horizontal line). Forecasting error was however large for the periods 06/02/2018 to 01/04/2018 and 25/062019 to 15/07/2019. These periods are indicated with the red dashed vertical lines. 

\begin{table}[H]
\caption{Evaluation metrics values for GLMNET, RF, SVR with linear Kernel, and Stacking ensemble algorithms}
\renewcommand{\arraystretch}{1.5}
	\centering	
\resizebox{\columnwidth}{!}{
		\begin{tabular}{>{\bfseries}c*{14}{c}} 
			\toprule
\multirow{2}{*}{\bfseries }&
\multicolumn{3}{c}{\bfseries MAPE (\%)}  &
\multicolumn{3}{c}{\bfseries RMSE (USD)} &
\multicolumn{3}{c}{\bfseries MAE (USD)}  &
\multicolumn{3}{c}{\bfseries R-Squared}
\\
\cmidrule(lr){3-4}
\cmidrule(lr){6-7} 
\cmidrule(lr){9-10} 
\cmidrule(lr){12-13} 
&& \textbf{Testing} & \textbf{Training}  
&& \textbf{Testing} & \textbf{Training} 
&& \textbf{Testing} & \textbf{Training} 
&& \textbf{Testing} & \textbf{Training}
			\\ \hline
GLMNET && 0.0194 & 0.5474 && 173.2304 & 78.3123 && 138.1717 & 31.0918 && 0.9966 & 0.9994      
\\
RF && 0.0548 & {\bf 0.0063} && 398.6882 & {\bf 44.0983} && 305.1938 & {\bf 10.2114} && 0.9835 & {\bf 0.9997}          
\\
SVR (linear) && 0.0209 & 3.2798 && 160.4642 & 139.4931 && 130.5756 & 121.4614 && 0.9960 & 0.9973          
\\
Stacking && {\bf 0.0191} & 0.9964 && {\bf 15.5331} & 76.3510 && {\bf 124.5508} & 48.7798 && {\bf 0.9967} & 0.9993         
\\
			\bottomrule
		\end{tabular}
	}
\label{evaluation_metrics_point}
\end{table}

\section{Conclusion}
\label{conclusion}
\noindent
In the existence of high volatility of Bitcoin price, an accurate and reliable forecasting models for Bitcoin price is very important for investors and market players. 
\\
Three machine learning models (generalized linear model via penalized maximum likelihood, random forest, support vector regression with linear kernel) were used to predict the price of bitcoin in the midst of price uncertainties. The construction of a stacking ensemble model using generalized linear model via penalized maximum likelihood, random forest as the base learners and support vector regression with linear kernel as the meta-learner reduced the prediction error for the three machine learning models, which was already low to begin with. Clearly, the stacking ensemble was functional in fine-tuning a model to attain a nearly perfect prediction. 
\\
The performance metrics (mean absolute percentage error, root mean square error, mean absolute error, and coefficient of determination) showed that the stacking ensemble model was the optimal model for predicting the testing data. However, the result is not to conclude that, the stacking ensemble model is superior to the other models; the performance of a model under separate states should be studied and understood. By employing machine learning techniques, the closing price of Bitcoins has been forecasted. Even though, the price of Bitcoin is very volatile, machine learning models were able to accurately forecast the price of Bitcoin.

\section*{Conflict of Interest}
The author declare that there are no conflicts of interest regarding the publication of this paper.

\section*{Data Availability}
Data for this work are available from the author upon request.

\bibliographystyle{model1-num-names}
\bibliography{sample.bib}

\end{document}